\documentclass[pre,twocolumn,superscriptaddress]{revtex4}

\usepackage{amssymb,amsbsy}
\usepackage{graphicx}
\usepackage{rotating}
\usepackage{dcolumn}
\usepackage{amsmath}

\newcommand{\e}{\mathrm{e}}
\newcommand{\Tr}{\mathop{\mathrm{Tr}}}
\newcommand{\Ord}{\mathrm{O}}

\newcommand{\av}[1]{\langle #1 \rangle}

\newcommand{\etal}{\textit{et al.}}
\newcommand{\mat}{\textbf}

\begin{document}
\title{Balance in signed networks}

\author{Alec Kirkley}
\affiliation{Department of Physics, University of Michigan, Ann Arbor, Michigan 48109, USA}

\author{George T. Cantwell}
\affiliation{Department of Physics, University of Michigan, Ann Arbor, Michigan 48109, USA}

\author{M. E. J. Newman}
\affiliation{Department of Physics, University of Michigan, Ann Arbor, Michigan 48109, USA}
\affiliation{Center for the Study of Complex Systems, University of Michigan, Ann Arbor, Michigan 48109, USA}

\begin{abstract}
  We consider signed networks in which connections or edges can be either positive (friendship, trust, alliance) or negative (dislike, distrust, conflict).  Early literature in graph theory theorized that such networks should display ``structural balance,'' meaning that certain configurations of positive and negative edges are favored and others are disfavored.  Here we propose two measures of balance in signed networks based on the established notions of weak and strong balance, and compare their performance on a range of tasks with each other and with previously proposed measures.  In particular, we ask whether real-world signed networks are significantly balanced by these measures compared to an appropriate null model, finding that indeed they are, by all the measures studied.  We also test our ability to predict unknown signs in otherwise known networks by maximizing balance.  In a series of cross-validation tests we find that our measures are able to predict signs substantially better than chance.
\end{abstract}
\maketitle

\section{Introduction}
Networks are used as an abstract representation of the topology of complex systems in many branches of science.  Examples include social networks of friendship or acquaintance between individuals, communication networks such as the Internet or telephone networks, infrastructure networks such as transportation routes, power grids, or pipelines, and information networks such as the World Wide Web or citation networks~\cite{Newman18c}.

In its simplest form, a network consists of a collection of nodes joined together in pairs by edges, but many networks have additional features as well.  The edges may be directed or weighted; either the nodes or edges may have types, categories, or labels of some kind; nodes may have positions in space; edges may have lengths or capacities, and so forth.  In this paper we consider one case of particular interest, that of \textit{signed networks}, meaning networks in which the edges are either positive or negative~\cite{Newman18c,Harary53,WF94}.  The most common example is a social network that represents patterns of both amity and enmity among a group of individuals: positive edges represent friendship, negative ones animosity.

Studies of signed networks go back at least to the classic work of Harary in the 1950s, who argued, largely on formal rather than empirical grounds, that certain patterns of signs should be more common than others---the enemy of my enemy should be my friend, for example~\cite{Harary53}.  Networks that display such regularities are said to be \textit{structurally balanced}, or just \textit{balanced} for short.  A natural question to ask is whether real signed networks are in fact balanced.  Despite a considerable amount of research on this issue, however, the jury is still out.  Some researchers have claimed that real networks are balanced, at least partially, while others have claimed that they are not~\cite{facchetti2011,estrada2014social,estrada2014walk}.

There are two primary reasons for the disagreement.  First, there is more than one proposed definition of structural balance in networks.  Cartwright and Harary~\cite{cartwright1956structural} proposed that a network is balanced if all closed loops in the network contain an even number of negative edges.  This condition, which we will refer to as \textit{strong balance}, is a stringent one that is rarely if ever completely satisfied in real networks.  As we will see, however, one can define measures of partial balance that quantify how close a network comes to Cartwright and Harary's ideal.

Strong balance is an attractive formulation in part because of a theorem due to Harary~\cite{Harary53}, which says that any network displaying perfect strong balance is \textit{clusterable}, meaning its nodes can be divided into some number of disjoint sets such that all edges within sets are positive and all edges between sets are negative.  Thus strong balance provides a possible theoretical basis for insularity or cliquishness in social networks: if networks naturally display strong balance, then they also naturally divide into communities such that people like members of their own community and dislike members of other communities.

While strong balance is a sufficient condition for clusterability, however, it turns out that it is not a necessary one, as shown by Davis~\cite{Davis67}, who demonstrated that for a network to be clusterable in the sense above, one requires only a lesser form of structural balance, namely that there be no closed loops in the network with exactly one negative edge.  We will refer to this condition as \textit{weak balance}.  Weakly balanced networks are a superset of strongly balanced ones---every strongly balanced network is necessarily also weakly balanced---but weak balance alone is enough to explain insularity in networks and division into antagonistic communities.

Alternatively, causality might run in the opposite direction: if a population is intrinsically divided into two or more antagonistic factions---Montagues and Capulets, Roundheads and Cavaliers, Hatfields and McCoys---then by definition the resulting network will be balanced.  Indeed, if there are exactly two factions then the network will be strongly balanced, since every closed loop must traverse negative edges between the factions an even number of times.  If there are three or more factions then the network will, in general, be only weakly balanced.

Thus we have two competing notions of what it means for a network to be balanced.  It is in part the lack of consensus about which of the two to adopt that makes it hard to say whether real networks are in fact balanced or not.

The second reason for the lack of agreement is that in order to say whether a network is balanced we need to specify the scale on which balance is to be assessed.  Even if we can agree on a measure of balance, how do we know whether the observed level is high or low?  A natural approach is to compare the level to what we would expect on the basis of chance, i.e.,~to the level in some kind of null model, but it is by no means universally agreed what form such a null model should take.

In this paper we do several things.  First, we consider a number of possible measures of both strong and weak balance.  Some of the measures we discuss have been proposed previously; some we propose here for the first time.  Second, we consider possible null models against which to compare levels of balance, choosing one we believe to be appropriate for the questions we are interested in.  Third, we use our measures and our null model to quantify structural balance in real-world signed networks, finding that the networks we consider are indeed significantly more balanced, at least according to some measures, than we would expect on the basis of chance.

The presence of structural balance in networks is interesting in its own right, for the hints it gives us about the growth and function of social networks.  But we can also use our knowledge of balance to perform other tasks.  As an example, we demonstrate how it can be used to make predictions of the signs of unobserved edges.  By simply assigning edges the choice of sign that makes the overall network most balanced, we show that we can predict the correct value of missing edge signs in test networks substantially better than chance.  As a corollary, this also gives us some insight about which are the best measures of balance: all of the measures we consider perform well in the sign prediction task, but the measure based the weak notion of balance appears to perform somewhat better, perhaps indicating that weak balance is a more correct description of the behavior of real-world networks than strong balance.

There has been a significant amount of previous work to define and study structural balance in signed networks, including methods and metrics motivated by spin glasses~\cite{facchetti2011,marvel2009energy}, dynamical systems~\cite{altafini2012dynamics,zheng2015social}, and spectral methods~\cite{acharya1980spectral,anchuri2012communities,kunegis2010spectral}, as well as walk-based approaches~\cite{harary1959measurement,norman1972derivation}, of which our own proposed methods can be considered an example.  Several recently proposed approaches share some features with our methods~\cite{chiang2011exploiting,chiang2014prediction,estrada2014walk,Singh2017}, although there are some crucial differences as well.  Perhaps the approach most similar to ours is that of Singh and Adhikari~\cite{Singh2017}, who propose a measure of balance motivated by the notion of strong balance that accounts (as ours also does) for the lesser effect of long loops on social tension.  We propose two similar measures, one for strong balance and one for weak, though with a different choice of weighting for short and long loops.  Another important difference between our work and that of Singh and Adhikari lies in the choice of null model, for which they use ensembles of networks where positive, negative, and non-edges are placed randomly.  In contrast, in our work we randomize only the signs of the edges and not their positions, which we argue is essential for proper quantification of statistically significant balance in networks.

\section{Quantifying balance}
\label{sec:defining}
Real-world signed networks are rarely, if ever, perfectly balanced, so to study balance in such networks we need a way to quantify exactly how balanced they are.  Following previous authors, we consider measures that quantify the number of closed loops in a network that violate either the strong or the weak notion of balance, meaning respectively that they have either an odd number of negative edges (strong balance) or exactly one negative edge (weak balance).

This alone, however, is not enough to define a practical measure because of another feature of networks, that the number of closed loops of a given length increases rapidly with length.  If one were simply to count closed loops, the count would be dominated by the longest loops in the network solely because they are more numerous.  It seems unlikely, however, that long loops play much of a role in real-world issues of balance.  Few people really care if a friend of a friend of a friend is an enemy or not.  Realistically, we expect that it is the short loops, not the long ones, that dominate network balance.  The second defining feature of the measures we consider, therefore, is that they weight short loops more heavily than long ones.

\subsection{Balance measures}
\label{sec:measures}
Consider an undirected signed graph or network~$G$.  A \emph{closed walk} in a such network is any path that begins and ends at the same node, and a \emph{simple cycle} is a closed walk that does not visit any node twice, other than the start/end node, which is visited exactly twice.  The strong definition of balance then says that $G$ is a balanced network if, and only if, every simple cycle in $G$ has an even number of negative signs.  The weak definition of balance, by contrast, says that a network is balanced if, and only if, it contains no simple cycles with exactly one negative edge (meaning that any other number is fine).  We can also say that individual cycles are strongly or weakly balanced by the same criteria.

We can use these ideas to define a measure~$B(z)$ of the level of imbalance in a network thus:
\begin{equation}
B(z) = \sum_{k=1}^\infty {I_k\over z^k},
\label{eq:Bmetric}
\end{equation}
where $I_k$ is the number of imbalanced simple cycles of length~$k$ and~$z>1$ is a free parameter.  This measure takes the form of a weighted count of imbalanced cycles in which longer cycles get downweighted by a geometric factor~$z^k$.  Note that the sum in~\eqref{eq:Bmetric} could in principle start at $k=2$ without changing the value of~$B(z)$, since there are no cycles of length one, but it will be convenient for subsequent developments to start at $k=1$.

We can define a measure of this type for either the weak or strong notion of balance.  Let us look first at the weak version, meaning that $I_k$ will be the number of simple cycles of length~$k$ that contain exactly one negative edge.  An immediate problem we encounter with applying this measure is the difficulty of making practical estimates of the number of simple cycles of a given length in an arbitrary network.  There is no elementary analytic approach for counting cycles, and numerical methods are hampered by the very rapid increase of $I_k$ with~$k$, which makes exhaustive enumeration of cycles possible only for small~$k$ and small networks.  Instead, therefore, we approximate the number of simple cycles by the number of closed walks, which is relatively straightforward to compute.  To count the number of weakly imbalanced closed walks of length~$k$, we remove all the negatives edges from the network and then look at the number of walks of length $k-1$ between the (former) endpoints of those edges.  Reinserting the negative edges again then closes the walks, creating loops of length exactly~$k$, each with exactly one negative edge.

Substituting closed walks for simple cycles is a good approximation when the cycles are short.  Indeed, for cycles of length three it is exact: closed walks and simple cycles are the same thing for length three.  As the length increases the approximation gets worse~\cite{latora2017complex}, but in practice this may not matter very much.  The imbalance metric of Eq.~\eqref{eq:Bmetric} discounts long loops, so the fact that our count is only approximate may not make much difference.

To put the developments in mathematical terms, let us denote the structure of our network by two adjacency matrices~$\mat{P}$ and~$\mat{N}$, for the positive and negative edges respectively.  Thus, matrix~$\mat{P}$ has elements~$P_{ij}=1$ if nodes $i$ and~$j$ are connected by a positive edge and 0 otherwise, and similarly $N_{ij}=1$ if $i$ and $j$ are connected by a negative edge and 0 otherwise.  Then our imbalance measure, which we will denote~$B_W(z)$ with subscript~$W$ to indicate weak balance, is given~by
\begin{equation}
B_W(z) = \tfrac12 \sum_{ij} N_{ij} \sum_{k=1}^{\infty} {1\over z^k}
         \bigl[ \mat{P}^{k-1} \bigr]_{ji}
       = \tfrac12 \Tr \bigl[ \mat{N} (z\mat{I} - \mat{P})^{-1}],
\label{eq:balance1}
\end{equation}
the factor of~$\frac12$ compensating for the fact that the sum counts each loop twice, once in each direction.

In fact, it will be convenient to introduce a rescaled parameter~$\alpha = z/\lambda_P$, where $\lambda_P$ is the leading (most positive) eigenvalue of~$\mat{P}$.  For $\alpha>1$ this ensures that the sum in~\eqref{eq:balance1} will converge, and we can write
\begin{equation}
B_W(\alpha) = \tfrac12 \Tr \bigl[ \mat{N}
              (\alpha\lambda_P\mat{I} - \mat{P})^{-1} \bigr].
\label{eq:bweak}
\end{equation}

Another way to interpret the parameter~$\alpha$ is to write $\alpha^{-k} = \e^{-k/k_0}$, where $k_0 = 1/\ln\alpha$ is a ``decay length'' that determines the length scale on which the contributions from longer walks are discounted.  Thus, for example, if we choose $\alpha=2$, we have $k_0 = 1/\ln2 \simeq 1.44\ldots$, and three such decay lengths give us a 95\% decay at distance a little greater than~4.

An analogous measure $B_S(\alpha)$ can be defined for the strong notion of balance.  Again we approximate the number of imbalanced simple cycles by the number of closed walks, which we can calculate as follows.  Consider the matrix $\mat{P}-\mat{N}$, which has elements $+1$ for positive edges, $-1$~for negative edges, and 0 otherwise.  The $k$th power of this matrix counts walks of length~$k$, times $+1$ if they contain an even number of minus signs and $-1$ if odd.  Thus the diagonal term $[(\mat{P}-\mat{N})^k]_{ii}$ is equal to the number of balanced closed walks starting and ending at node~$i$ minus the number of imbalanced ones.  Summing over all~$i$, we then have~\cite{latora2017complex}
\begin{equation}
B_k - I_k = {1\over 2k} \Tr \bigl[ (\mat{P}-\mat{N})^k \bigr],
\label{eq:difference}
\end{equation}
where $B_k$ and $I_k$ are the total number of balanced and imbalanced closed walks.  The initial factor of $\frac12$ again compensates for the fact that we count each loop in both directions, and the factor of $1/k$ compensates for the fact that each loop is counted repeatedly starting from each of the $k$ points along its length.

Conversely, consider the matrix $\mat{P}+\mat{N}$, which is simply the adjacency matrix of the complete network, ignoring signs---every edge, positive or negative, is represent by a $+1$ in this matrix.  The total number of closed walks of length~$k$, both balanced and imbalanced, is given by
\begin{equation}
B_k + I_k = {1\over 2k} \Tr \bigl[ (\mat{P}+\mat{N})^k \bigr].
\label{eq:sum}
\end{equation}
Subtracting~\eqref{eq:difference} from~\eqref{eq:sum} and dividing by~2, we get an expression for the number of imbalanced loops:
\begin{equation}
I_k = {1\over 4k} \Tr \bigl[ (\mat{P}+\mat{N})^k \bigr]
      - {1\over 4k} \Tr \bigl[ (\mat{P}-\mat{N})^k \bigr].
\end{equation}
Substituting this into Eq.~\eqref{eq:Bmetric} then gives us our measure of strong imbalance:
\begin{equation}
B_S(z) = \tfrac14 \sum_{k=1}^\infty
         {1\over k z^k} \Tr \bigl[ (\mat{P}+\mat{N})^k \bigr]
         - \tfrac14 \sum_{k=1}^\infty {1\over k z^k}
         \Tr \bigl[ (\mat{P}-\mat{N})^k \bigr].
\label{eq:bstrong1}
\end{equation}
Making use of the matrix identity
\begin{equation}
\sum_{k=1}^\infty {\Tr \mat{M}^k\over k} = \log \det(\mat{I}-\mat{M}),
\end{equation}
this can also be written as
\begin{equation}
B_S(z) = \tfrac14 \log {\det[z\mat{I} - (\mat{P}-\mat{N})]\over
                        \det[z\mat{I} - (\mat{P}+\mat{N})]},
\label{eq:bstrong2}
\end{equation}
which is valid whenever the sums in~\eqref{eq:bstrong1} converge.  As with $B_W(z)$ it is convenient to reparametrize this expression in terms of $\alpha = z/\lambda^*$, where $\lambda^*$ is the larger of the leading eigenvalues of $\mat{P}+\mat{N}$ and $\mat{P}-\mat{N}$, so that
\begin{equation}
B_S(z) = \tfrac14 \log
  {\det[\alpha\lambda^*\mat{I} - (\mat{P}-\mat{N})]\over
   \det[\alpha\lambda^*\mat{I} - (\mat{P}+\mat{N})]},
\label{eq:bstrong3}
\end{equation}
which ensures convergence of the sums when $\alpha>1$.

\subsection{Previous measures of network balance}
A number of previous researchers have also proposed measures of structural balance in networks.  Estrada and Benzi~\cite{estrada2014walk} (henceforth EB) define a measure
\begin{equation}
B_{\textrm{EB}} = {1-K\over 1+K},
\label{eq:estrada}
\end{equation}
where
\begin{equation}
 K= {\sum_k \Tr\bigl[ (\mat{P}-\mat{N})^k \bigr]/k!
  \over \sum_k \Tr\bigl[ (\mat{P}+\mat{N})^k \bigr]/k!}.
\end{equation}
The quantity~$K$ is in some ways analogous to our measure of strong imbalance, Eq.~\eqref{eq:bstrong2}, but it downweights longer loops by a larger factor~$1/k!$, compared to the geometric factor~$1/z^k$ that we employ.  This results in some elegant mathematical expressions but has the disadvantage that there is no way to set the length scale on which loops are discounted.  EB~also define their measure not by $K$ itself but by the formula~\eqref{eq:estrada}, which can be interpreted as a ratio of weighted counts of unbalanced and balanced loops.

Singh and Adhikari~\cite{Singh2017} (henceforth SA), in considering the measure of~EB, object to the weight factor~$1/k!$ and propose instead to use a geometric factor as we do, defining a measure
\begin{equation}
B_{\textrm{SA}}(z) = {\sum_k \Tr \bigl[ (\mat{P}-\mat{N})^k \bigr]/z^k\over
                    \sum_k \Tr \bigl[ (\mat{P}+\mat{N})^k \bigr]/z^k}.
\end{equation}
This is again somewhat analogous to Eq.~\eqref{eq:bstrong2}, though it is not directly based on the actual number of imbalanced loops, and moreover appears to neglect the factor of $1/k$ that accounts for the $k$ possible starting points around a loop of length~$k$.

In this paper we compare the performance of the four measures discussed here, our own measures $B_W$ and $B_S$ and the measures of EB and SA, on a number of problems concerning balance in networks.

\subsection{Null models}
\label{sec:null}
As discussed in the introduction, measures of imbalance are difficult to employ on their own because we lack a scale on which to calibrate their values.  If we calculate a value of, say, $B_W=0.5$ for a particular network how do we know if that value is large or small?  One way to answer this question is to compare our numbers with values calculated in an appropriate null model.

The broader question we are addressing in calculating measures of balance is whether the arrangement of positive and negative edges within a network is somehow special, different from what we would expect on the basis of chance.  Since our focus is on the arrangement of signs within the larger network, and not on the arrangement of edges per se, the natural null model to consider is one in which the signs in a network are randomized while keeping the locations of the edges fixed.  In the particular null model we consider here, we also keep the overall number of positive and negative signs fixed, to make the randomized networks more directly comparable with the original.

This null model or ones similar to it have been used in a number of previous works~\cite{GKRT04,Mouttapa04,XYG09}, but it is not the only possible choice~\cite{Singh2017,Feng18}.  Singh and Adhikari~\cite{Singh2017}, for example, employ a null model in which both the signs and the positions of the edges are randomized.  This results in networks whose structure, in terms of edge placement, is very different from that of the original network, which makes it difficult to know how much of any observed difference in balance is due to the pattern of signs and how much to the edge positions.  The null model we employ avoids this difficulty by randomizing the signs only.

Arguably, in many real-world situations---coworkers in an office, for instance, or children in a school class---one indeed has no choice about who one interacts with, so that the positions of the network edges are fixed.  The only degree of freedom is the nature of the interactions, whether they will be friendly or antagonistic.  A model that fixes the edge positions but varies their signs is thus a natural choice in such cases.

\section{Example applications}
\label{sec:applications}
As examples of the techniques introduced here, we consider their application to two data sets, one from the field of international relations, representing positive and negative ties between countries~\cite{izmirlioglu2017correlates}, and the other from sociology, representing ties between a group of university freshmen~\cite{van1999friendship}.  For both data sets we use our measures to quantify structural balance, and for the international relations data we also test our ability to make predictions of the signs of unobserved edges.  

The international relations data set contains many details of inter-country interactions over a period of several decades, but here we focus on two aspects in particular: alliances and wars.  We construct a set of signed networks, one for each year in the 70-year period from 1938--2008, in which nodes represent countries and two countries are connected by a positive tie if they have a formal alliance in that year and a negative tie if there is a militarized dispute between them.  In the rare cases in which countries have both an alliance and a war in the same year we take the corresponding edge to be negative.  (The same methodology was used previously in~\cite{lerner2016structural}.)  Only countries for which we have data are included in our networks.  The number of nodes ranges from 25 to 155 with a median of~105, and the number of edges ranges from 46 to 1230 with a median of 615.  The signs of the edges are predominantly positive---most countries have good relations.  The fraction of negative edges ranges from 1.8\% to 45.1\% with a median of 5.5\%.  (The outliers with the largest number of negative edges all fall during the Second World War.  The median fraction of negative edges between 1940 and 1945 was 44\%.)

The university freshman data set describes relationships between a group of first-year students, all at the same university, and consists of networks collected at seven different time points.  At each time point the students were asked to rate their relationships with all other students in the group on a five-point scale of (1)~``best friend'', (2)~``friendship'', (3)~``friendly relationship'', (4)~``neutral relationship'', or (5)~``troubled relationship''.  Students could also say they did not know the person in question.  Further discussion of the scale can be found in~\cite{van1999friendship}.  We construct a set of signed networks, one for each time point, in which two students are connected by a positive tie if each rates the other as a 3 or lower, and a negative tie if one or both rates the other as a~5.  Neutral relationships are not represented in the network, which means that there is no difference in our representation between having a neutral relationship and having no relationship at all.  While this is not ideal, it seems like the best strategy given that there is no principled way to decide whether a neutral edge should be considered positive or negative.  Of the seven networks constructed in this way, we discard three because of sparse or missing data, leaving four that we analyze here.  The number of nodes in the networks is 34 at all time points and the number of edges ranges from 174 to 227 with a median of 225.5.  The fraction of negative edges ranges from 12\% to 14\% with a median of~13\%.

\subsection{Balance relative to the null model}
\label{sec:balance}
\begin{figure}
\centering
\includegraphics[width=8.3cm]{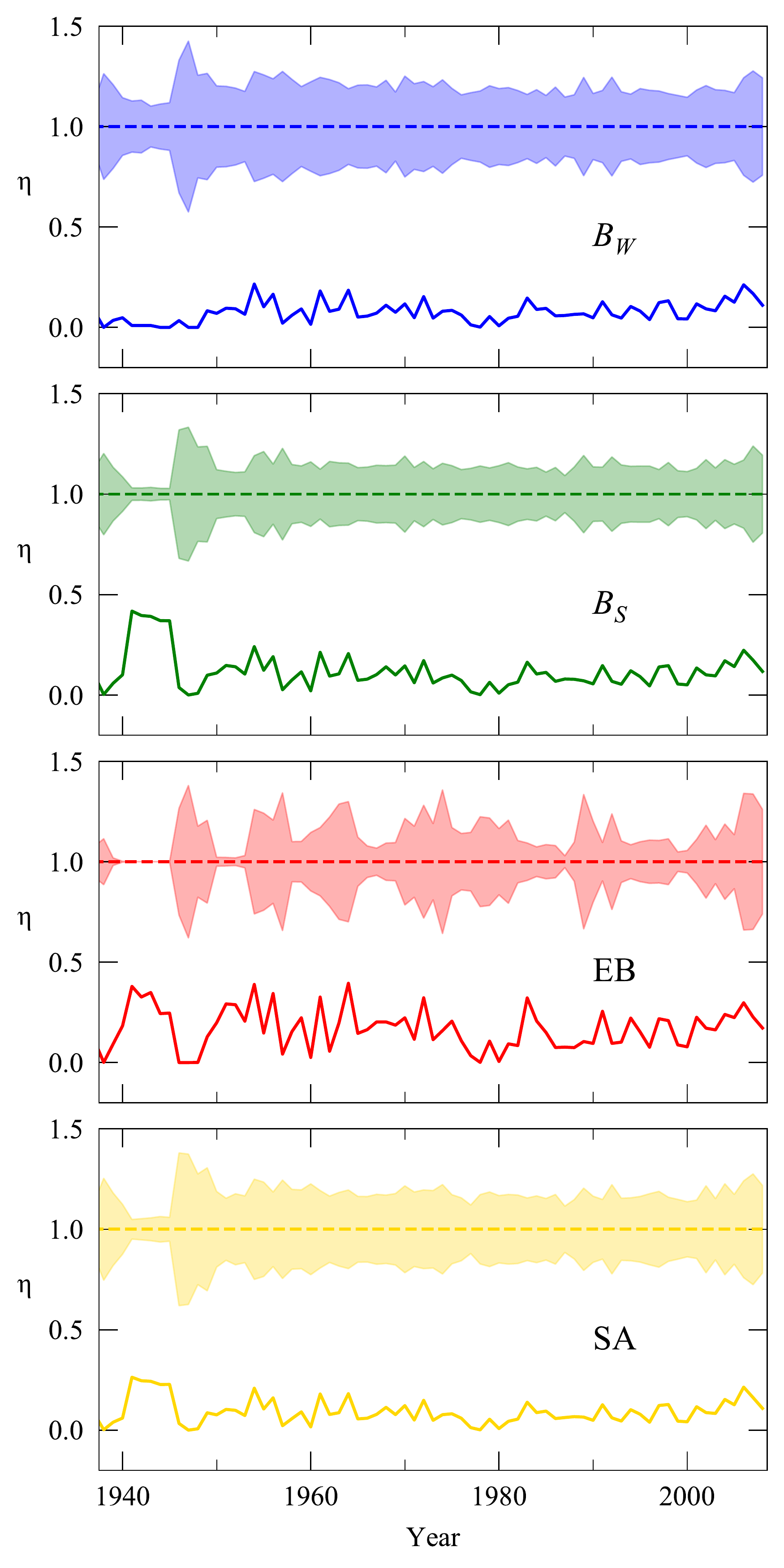}
\caption{Level of imbalance in the international relations networks for 1938--2008, as measured by the ratio~$\eta$ defined in Eq.~\eqref{eq:eta_imbalance}, for each of the four balance measures studied here.  The dotted lines indicate the null model mean, which falls at $\eta=1$ by definition, and the surrounding bands denote two standard deviations of the fluctuations around this mean.  The solid lines represent the values for the observed networks.  All networks are significantly less imbalanced than the null model by all four measures.}
\label{fig:imbalance_diplomacy}
\end{figure}

To quantify the level of balance in a network, we compute the ratio between the value~$B$ of each metric and the average value~$\av{B}$ of the same metric on a selection of randomized networks drawn from the null model described in Section~\ref{sec:null}:
\begin{equation}
\eta = {B \over\av{B}}.
\label{eq:eta_imbalance}
\end{equation}
Figure~\ref{fig:imbalance_diplomacy} shows the values of this ratio as a function of time for the international relations networks for the four balance metrics considered in this paper, along with an indication of the fluctuation of the results about the mean for the null model (the bands shown are two standard deviations).  As the figure shows, in each case actual imbalance values, for all measures, are far below what would be expected for the null model.  (An alternative way to represent the same results would be to compute a $z$-score, but we prefer the representation of Fig.~\ref{fig:imbalance_diplomacy} since it shows explicitly the size of the fluctuations in the null-model values.)

For this calculation the metrics $B_W$ and $B_S$ are both computed with a parameter value of~$\alpha=2$, as discussed on Section~\ref{sec:measures}, and we use the corresponding value for the parameter in the metric of Singh and Adhikari (SA)~\cite{Singh2017} as well.  (The metric of Estrada and Benzi (EB)~\cite{estrada2014walk} has no free parameters.)  We have also experimented with a range of alternative parameter values, but find that the results do not depend strongly on our choice.

\begin{figure}
\centering
\includegraphics[width=8.3cm]{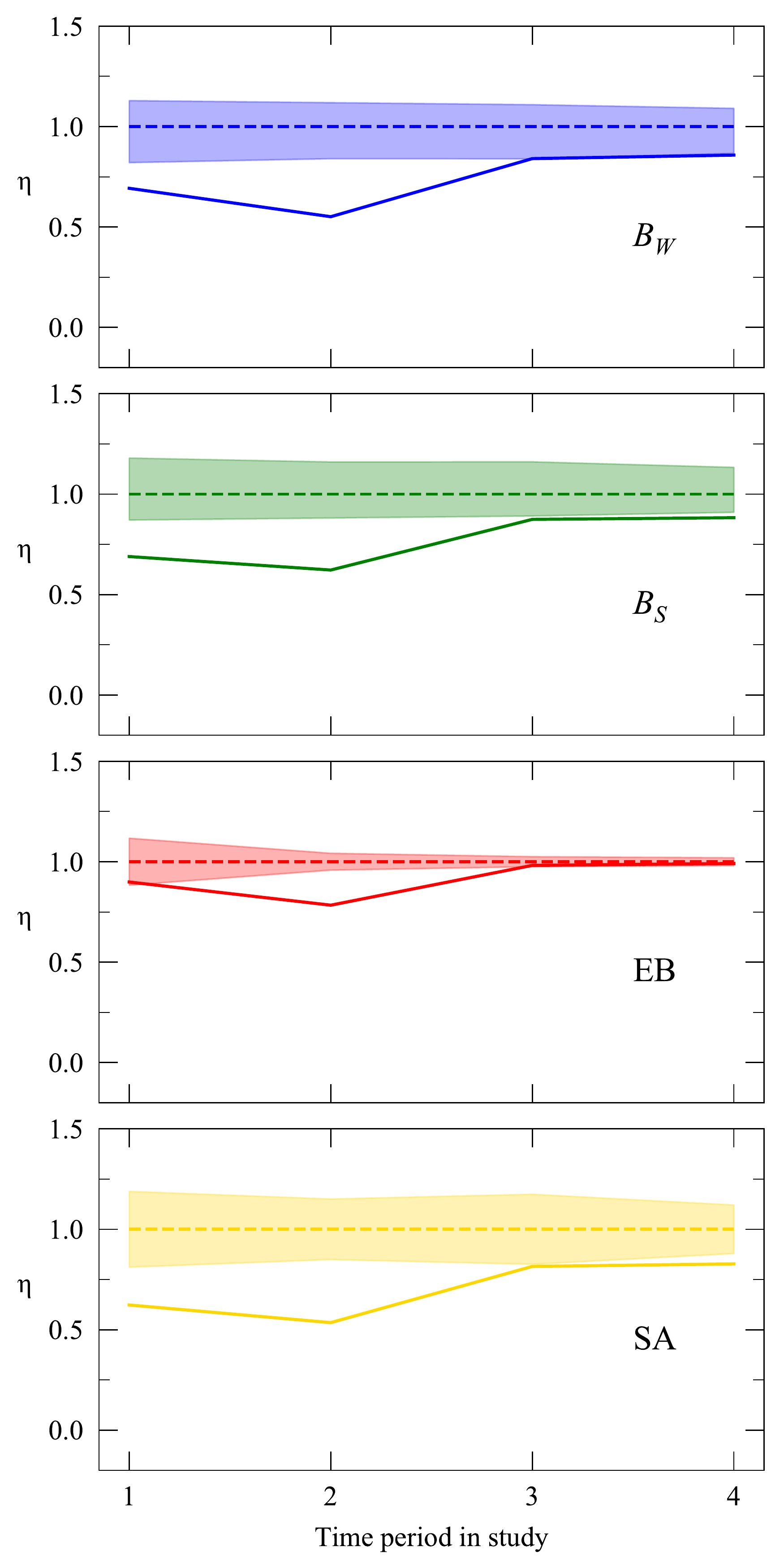}
\caption{Levels of imbalance in the university freshman networks.  The dotted lines indicate the null model mean and the surrounding bands denote two standard deviations of the fluctuations around this mean.  The solid lines represent the values for the observed networks.}
\label{fig:imbalance_school}
\end{figure}

Figure~\ref{fig:imbalance_school} shows results from the same experiment performed on the university freshman networks.

Our goals here are two-fold.  First, we wish to see if real networks are indeed unusually balanced.  Second, if they are balanced, we wish to see which of our notions of balance best captures that behavior (meaning which one gives the most statistically significant results).  As Fig.~\ref{fig:imbalance_diplomacy} shows, all four metrics give extremely low $\eta$ values relative to the null model, all of which would be statistically significant at the $p<0.05$ level in all years if we assume a normal distribution within the null model.  The most significant values occur during the World War II period, specifically between 1940 and 1945, and this effect is especially pronounced for the three metrics based on the strong notion of balance.  As mentioned in the introduction, strong balance is expected in cases where a network is divided into just two factions, which was the case during World War II.

Figure~\ref{fig:imbalance_school} for the university networks shows similar behavior, although the $\eta$ values are less extreme than those for the international relations networks.  This might be due to a lower level of factionalism for the students than for international relations, or to measurement error, or a combination of both.

The university data set also lends itself to being represented as a weighted, directed network, and one could consider generalizations of the methods presented here to such networks, although this is outside the scope of the present paper.

\subsection{Sign prediction}
\label{sec:singlesign}
Consider a situation in which we know the positions of the edges in a signed network, but we know the signs of only some of the edges.  The signs of the remaining edges are missing from our data, perhaps because they were not measured or recorded, or because our measurements are unreliable.  Guha~\etal~\cite{GKRT04}, in studies of trust in online communities, suggested that it should be possible, using the patterns of known signs, to make predictions about the unknown ones, and in recent years a number of authors have developed algorithms to do this~\cite{leskovec2010predicting,kunegis2010spectral,chiang2011exploiting,yang2012friend,chiang2014prediction,Singh2017}.

A natural approach is to start from the assumption that the network is balanced~\cite{leskovec2010predicting,Singh2017}.  Consider the simple case where the sign is missing from just a single edge in the network and our goal is to guess the value of that sign given all the others.  We assume that the best guess for the missing sign is the one that will make the network most balanced.  This leaves open the question of which metric we should use to quantify balance, which we address by performing a cross-validation study in which we artificially remove one sign from an otherwise complete network, then attempt to predict its value using each of our metrics in turn.  Repeating this process for every edge in the network, we measure the average success of our predictions for each metric.

This ``single edge'' prediction test is arguably unrealistic---in most real-world scenarios there will be more than one sign missing from any incomplete data set, a point that we discuss further in Section~\ref{sec:multiple}.  It is, nonetheless, a good starting point by virtue of being relatively computationally tractable for networks of the size considered here, which typically have a few hundred edges.  We can just calculate directly the value of each of our balance metrics for the two possible choices of each sign and take the choice that gives the higher balance.

For larger network sizes this brute-force approach becomes more computationally demanding, but with a little ingenuity the calculation can still be done.  The calculation of our metrics $B_W$ and $B_S$ relies on the computation of either a matrix inverse (for~$B_W$) or a matrix determinant (for~$B_S$), and there exist formulas that allow one to quickly recalculate inverses and determinants when only a few elements of a matrix are altered, as in this case.  Consider, for instance, the weak balance measure~$B_W$ defined in Eq.~\eqref{eq:balance1}.  The primary computational task in evaluating this measure is the calculation of the \textit{resolvent matrix} $\mat{R} = (z\mat{I} - \mat{P})^{-1}$.  We can speed up this calculation as follows.  First, we directly compute~$\mat{R}$ for the original network and use it to evaluate~$B_W$.  This is a relatively slow operation: computing the inverse of an $n\times n$ matrix takes $\Ord(n^3)$ time in a naive implementation, and modestly better in more complex schemes.  Then we consider in turn each edge in the network and compute the value of~$B_W$ when the sign of that edge is reversed.  Reversing the sign of an edge between nodes $i$ and~$j$ alters the values of $P_{ij}$ and~$P_{ji}$ by $\pm1$, a change that we can write in the low-rank form
\begin{equation}
\mat{P}' = \mat{P} \pm \mat{U}\mat{V},
\end{equation}
where $\mat{U}$ is an $n\times 2$ matrix with all elements zero except $U_{i1}=U_{j2}=1$, and $\mat{V}$ is a $2\times n$ matrix with all elements zero except $V_{1j}=V_{2i}=1$.  Then the \textit{Woodbury matrix identity}~\cite{Hager89} tells us that the new value of the resolvent~$\mat{R}' = (z\mat{I} - \mat{P}')^{-1}$ is given by
\begin{equation}
\mat{R}' = \mat{R} \pm \mat{R}\mat{U}(\mat{I} \mp \mat{V}\mat{R}\mat{U})^{-1}
           \mat{V}\mat{R},
\end{equation}
which requires only the trivial inversion of the $2\times 2$ matrix inside the brackets.  Evaluation of the matrix products $\mat{R}\mat{U}$ and $\mat{V}\mat{R}$ and evaluation of the $n^2$ elements of $\mat{R}'$ all take $\Ord(n^2)$ time, so the running time to calculate the new value of~$B_W$ is also~$\Ord(n^2)$, a substantial improvement on the $\Ord(n^3)$ time needed to calculate it from scratch.

Similarly for the strong balance measure~$B_S$ it is possible to evaluate the measure rapidly upon the change of single sign.  This measure, defined in Eq.~\eqref{eq:bstrong2}, involves the calculation of the determinant of the matrix~$\mat{A} = z\mat{I}-(\mat{P}-\mat{N})$, whose value changes upon the flip of a sign to
\begin{equation}
\mat{A}' = \mat{A} \pm 2\mat{U}\mat{V},
\end{equation}
where $\mat{U}$ and $\mat{V}$ are as previously defined.  (The determinant in the denominator of Eq.~\eqref{eq:bstrong2} does not change when a sign is flipped, so there is no need to recalculate it.)  Then the \textit{matrix determinant lemma}~\cite{Strang09} states that the new value of the determinant is related to the old one~by
\begin{equation}
\det(\mat{A} \pm 2\mat{U}\mat{V}) = \det(\mat{A})
  \det(\mat{I} \pm 2\mat{V}\!\mat{A}^{-1}\mat{U}).
\end{equation}
Once one has the inverse~$\mat{A}^{-1}$ this computation can be performed quickly.  The $2\times2$ matrix $\mat{I}\pm2\mat{V}\!\mat{A}^{-1}\mat{U}$ can be calculated in time~$\Ord(n^2)$ and its determinant in constant time, so again the overall calculation takes~$\Ord(n^2)$ time.  By contrast, calculating the determinant directly from scratch takes $\Ord(n^3)$ time (or slightly better using the fastest algorithms), so again we have a substantial improvement in speed over the direct calculation.  For the other balance metrics considered here (EB and SA) there are similar shortcuts that can speed up calculations for larger networks, although we will not use them here.

\begin{figure}
\centering
\includegraphics[width=\columnwidth]{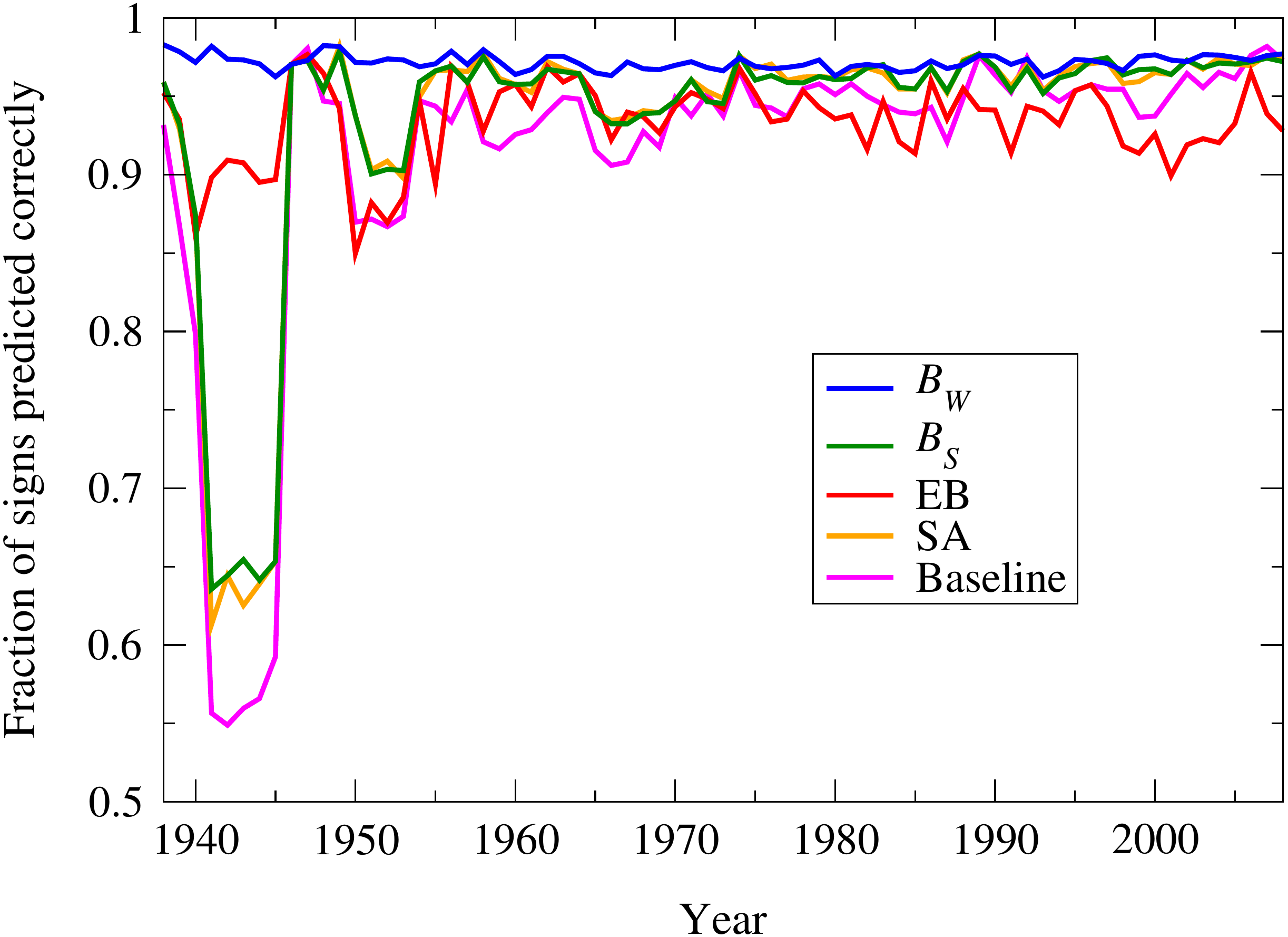}
\caption{Fraction of signs predicted correctly for each of the international relations networks in the single sign prediction task of Section~\ref{sec:singlesign}, using each of the four balance measures studied here.}
\label{fig:acc}
\end{figure}

Figure~\ref{fig:acc} shows the results of single-sign prediction calculations for our international relations networks as a function of time, for each of our four measures of balance.  The vertical axis in the figure measures the fraction of all signs predicted correctly, also known as the \textit{accuracy}.  By contrast with the results shown in Figs.~\ref{fig:imbalance_diplomacy} and~\ref{fig:imbalance_school}, performance on this task clearly varies between the different balance metrics, and in particular the measure~$B_W$ based on the weak notion of balance performs significantly better than any of the strong balance measures.

One must be a little careful about these results, however, because, as mentioned previously, positive edges outnumber negative ones by a wide margin in most cases.  This means that one can achieve quite high prediction accuracy simply by guessing that every edge is positive.  The magenta curve in Fig.~\ref{fig:acc} represents this baseline level of accuracy and it is against this curve that the others should be judged.  Thus, for example, the measure of EB, which gave generally good performance in Fig.~\ref{fig:imbalance_diplomacy}, performs least well in terms of sign prediction accuracy and in some cases is actually below the baseline estimate, particularly in the latter half of the data set.  Meanwhile, the weak balance measure~$B_W$ substantially outperforms the other measures and the baseline, and appears to give the best sign prediction performance of the measures considered.

\begin{figure}
\centering
\includegraphics[width=\columnwidth]{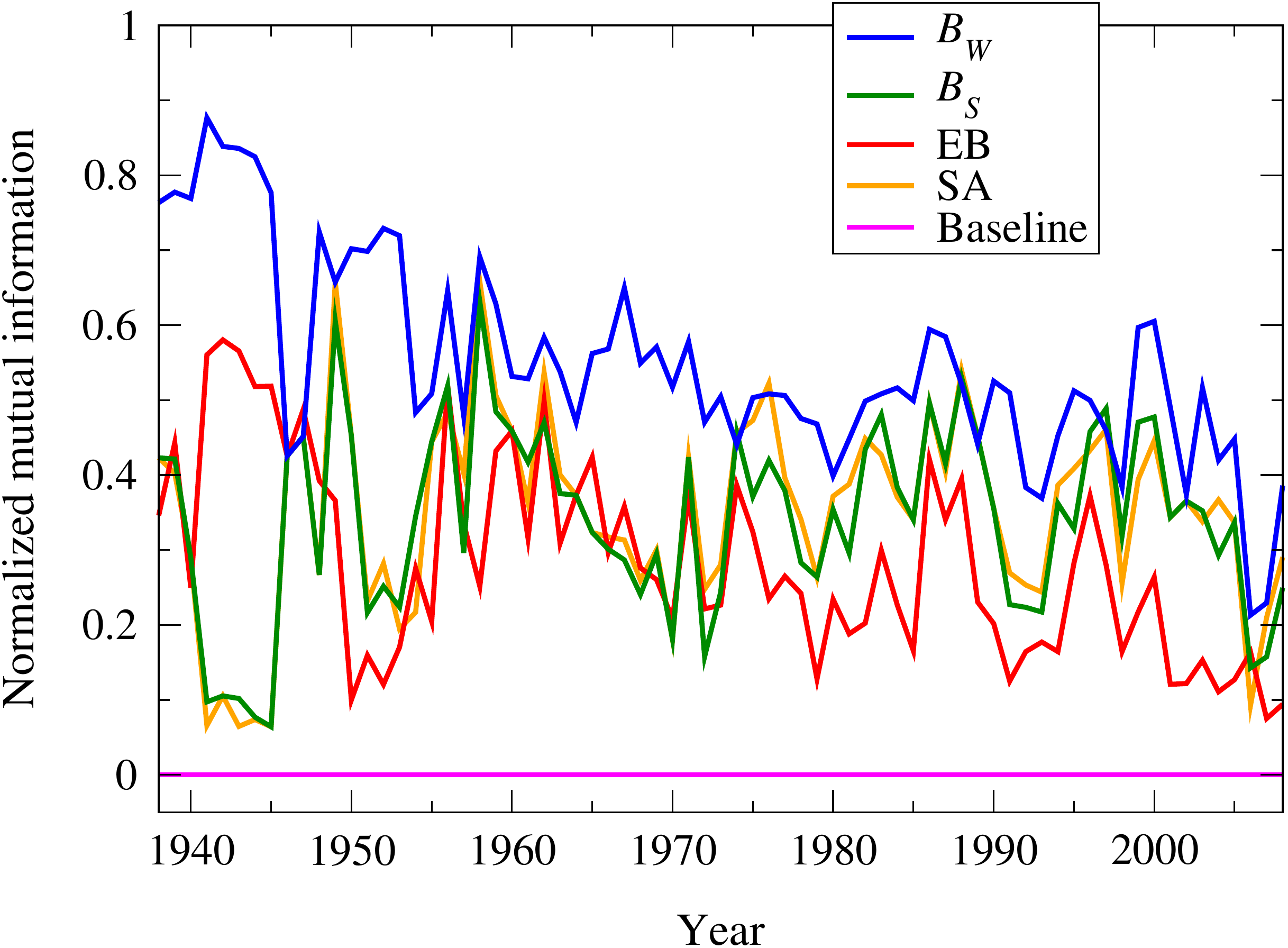}
\caption{Success in the single sign prediction task, as measured using normalized mutual information, for each of the four balance measures studied here.}
\label{fig:NMI}
\end{figure}

Figure~\ref{fig:NMI} shows an alternative measure of prediction performance, the normalized mutual information~\cite{DDDA05}.  Often used to quantify the success of community detection algorithms on networks, normalized mutual information (NMI) is an information theoretic measure that reflects the amount of information about the true signs of edges that is contained in the predicted signs.  If the predicted signs match the true signs exactly, the NMI is~1; if there is no correlation between true and predicted signs it is zero.

The (unnormalized) mutual information between true signs~$s_t$ and predicted signs~$s_p$ is defined as
\begin{equation}
I(s_t;s_p)\> = \sum_{\substack{s_t=\pm1\\s_p=\pm1}} P(s_t,s_p) \log
  \frac{P(s_t,s_p)}{P(s_t) P(s_p)}.
\label{eq:MI}
\end{equation}
The joint probabilities~$P(s_t=\pm1,s_p=\pm1)$ can be calculated straightforwardly by simply counting the fraction of times in our tests that each of the four possible configurations of the true and predicted signs occurs, and similarly for the marginal probabilities~$P(s_t=\pm1)$ and $P(s_p=\pm1)$.  The \emph{normalized} mutual information is then calculated by dividing the unnormalized value by the average of the entropy~$H(s)=-\sum_s P(s) \log P(s)$ of the two variables $s_t$ and~$s_p$~\cite{DDDA05}:
\begin{equation}
\text{NMI} = \frac{I(s_t;s_p)}{\tfrac12[H(s_t)+H(s_p)]}.
\end{equation}
This ensures that the normalized value falls between zero and one.

As shown in Fig.~\ref{fig:NMI}, the normalized mutual information for sign prediction using all four of our balance measures is better than the baseline estimate made by simply guessing that all edges have the majority positive sign---the latter automatically gets an NMI of zero, since it is completely uncorrelated with the true signs of the edges.  Again the weak balance measure~$B_W$ does best in most years, in some cases by a wide margin.

Comparing our results from this section with those on overall balance from Section~\ref{sec:balance}, we see something of a mixed picture. Overall balance appears to be similar for all metrics, aside from the era of the Second World War where the strong notion of balance seems to be favored.  Our sign prediction results, on the other hand, seem to give a clear edge to the weak notion of balance, even during the war years.

What we can say with some clarity is that these networks are more balanced than one would expect on the basis of chance, and one can use this fact to predict the signs of edges with good accuracy.

\subsection{Prediction of multiple edge signs}
\label{sec:multiple}
In the calculations of the previous section, we tested our ability to predict a single unknown sign in an otherwise known network.  This single sign prediction challenge has the advantage of being relatively computationally tractable, but, as we have argued, it is not entirely realistic.  In real-world data sets it is likely that many signs will be missing from our network simultaneously, not just one, and hence we need a way to predict multiple signs simultaneously.  We can approach the latter problem in a similar manner to single edge prediction, by selecting the combination of signs that gives the lowest imbalance, but the calculation rapidly becomes intractable as the number~$k$ of signs to be predicted becomes large, since there are $2^k$ different combinations of signs to test.

To get around this issue, we employ simulated annealing to optimize balance over sign configurations.  We perform a Markov chain Monte Carlo simulation in which we initially give random values to all of the unknown signs, then we repeatedly select one of them at random and consider flipping its value, from positive to negative or vice versa.  We can use any one of our imbalance metrics as an energy function and accept or reject sign flips using a standard Metropolis--Hastings acceptance probability with temperature~$T$.  We then lower the temperature from a high initial value~$T_0$ according to the exponential cooling schedule $T = T_0\,\e^{-t/\tau}$, where $t$ is the number of Monte Carlo steps performed so far and $\tau$ is the annealing time-scale.  The calculation ends when the state of the system stops changing and we take the final state to be our prediction of the unknown signs.

For the calculations presented here we use parameter values $T_0 = 0.1$ and $\tau = 10^4$ and run our calculations for $10^6$ Monte Carlo steps.  For each network studied, we remove varying fractions of the signs and then attempt to predict those removed, repeating the entire calculation 100 times for each fraction.  For the imbalance measures used in this paper the calculation can be sped up significantly by rapidly computing the new energy value upon the flip of a sign using the Woodbury or matrix determinant formulas again.  Here we focus specifically on the measures $B_W$ and~$B_S$.  Since these measures are constructed in an identical manner apart from the criteria they use for balanced loops, they give us an opportunity to perform an apples-to-apples comparison of strong and weak notions of balance, to see which gives better sign prediction.  Similar calculations would, however, certainly be possible for the EB and SA metrics considered in previous sections.

\begin{figure}
\centering
\includegraphics[width=\columnwidth]{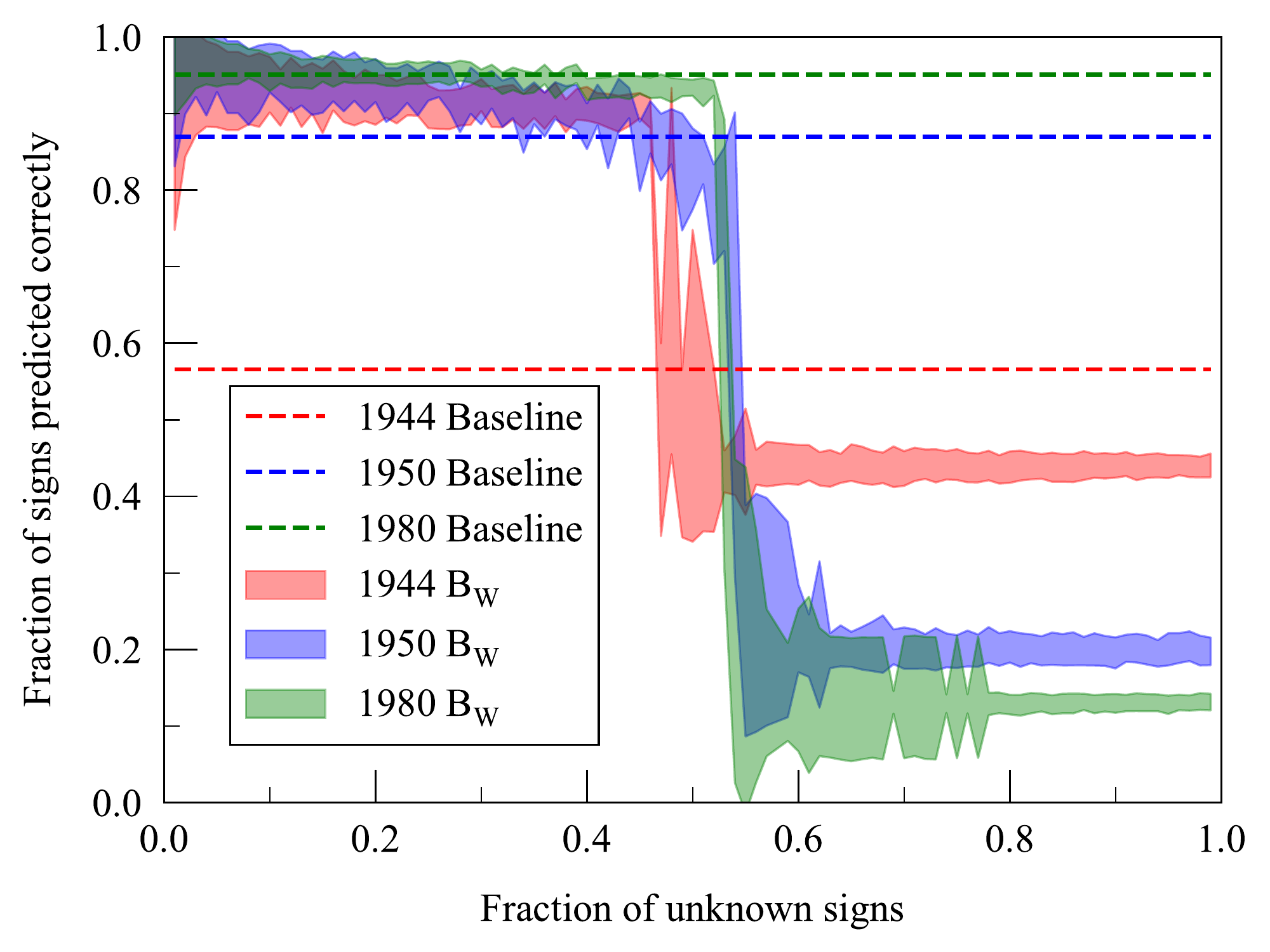}
\caption{Fraction of signs predicted correctly in the multiple sign prediction task using the weak balance measure~$B_W$, as a function of the fraction of unknown signs for the international relations networks in the years 1944, 1950, and 1980, along with baseline levels derived by simply assuming all signs to be positive.  Bands indicate $1\sigma$ errors calculated from the distribution of values over 100 randomized repetitions of the calculation.}
\label{fig:weak_acc}
\end{figure}

\begin{figure}
\centering
\includegraphics[width=\columnwidth]{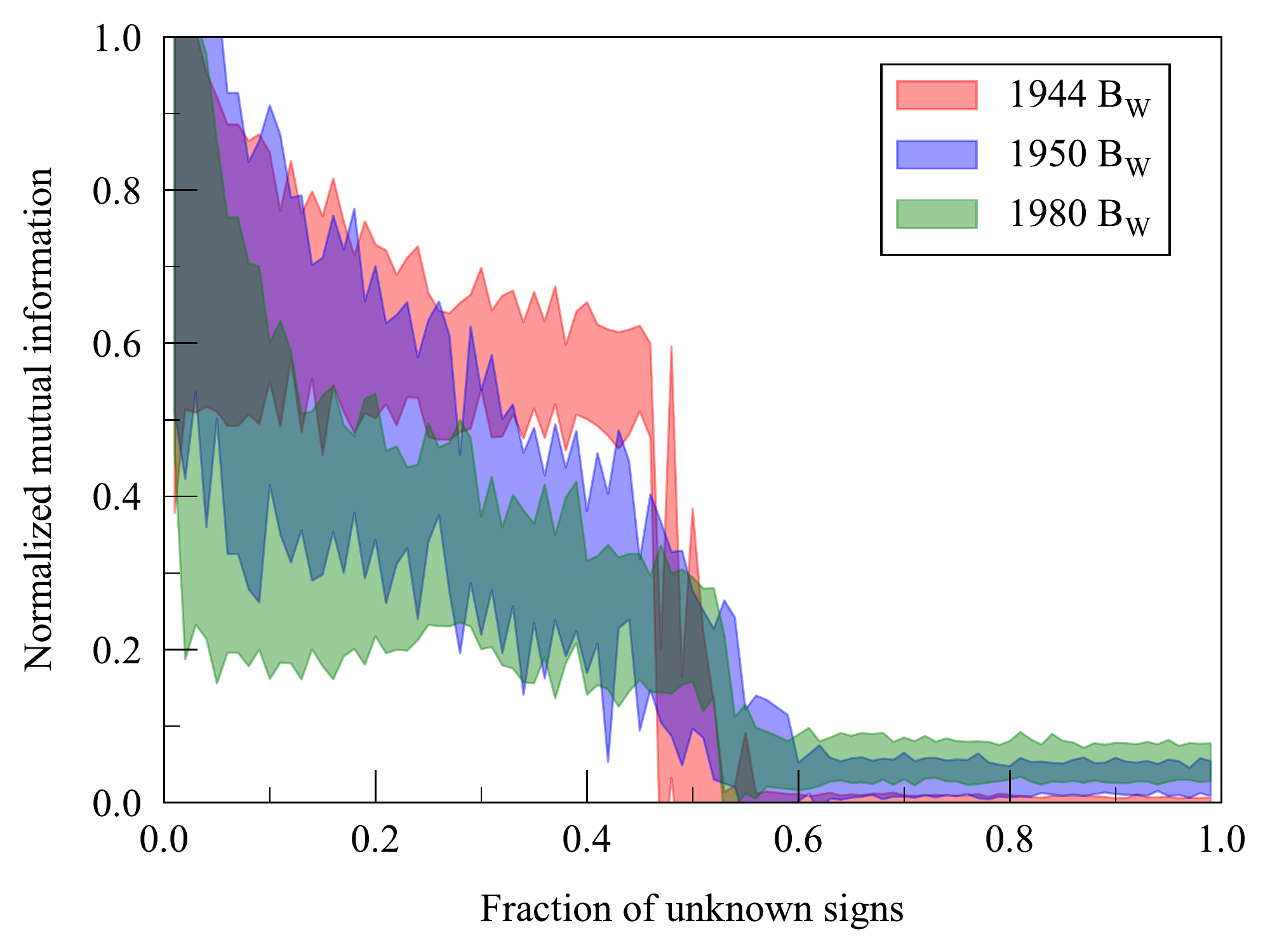}
\caption{Normalized mutual information for the multiple sign prediction task using the weak balance measure~$B_W$, as a function of the fraction of unknown signs for the international relations networks in the years 1944, 1950, and 1980.  The baseline level of normalized mutual information if we guess all signs to be positive is zero by definition.  Bands indicate $1\sigma$ errors calculated from the distribution of NMI values over 100 randomized repetitions of the calculation.}
\label{fig:weak_nmi}
\end{figure}

\begin{figure}
\centering
\includegraphics[width=\columnwidth]{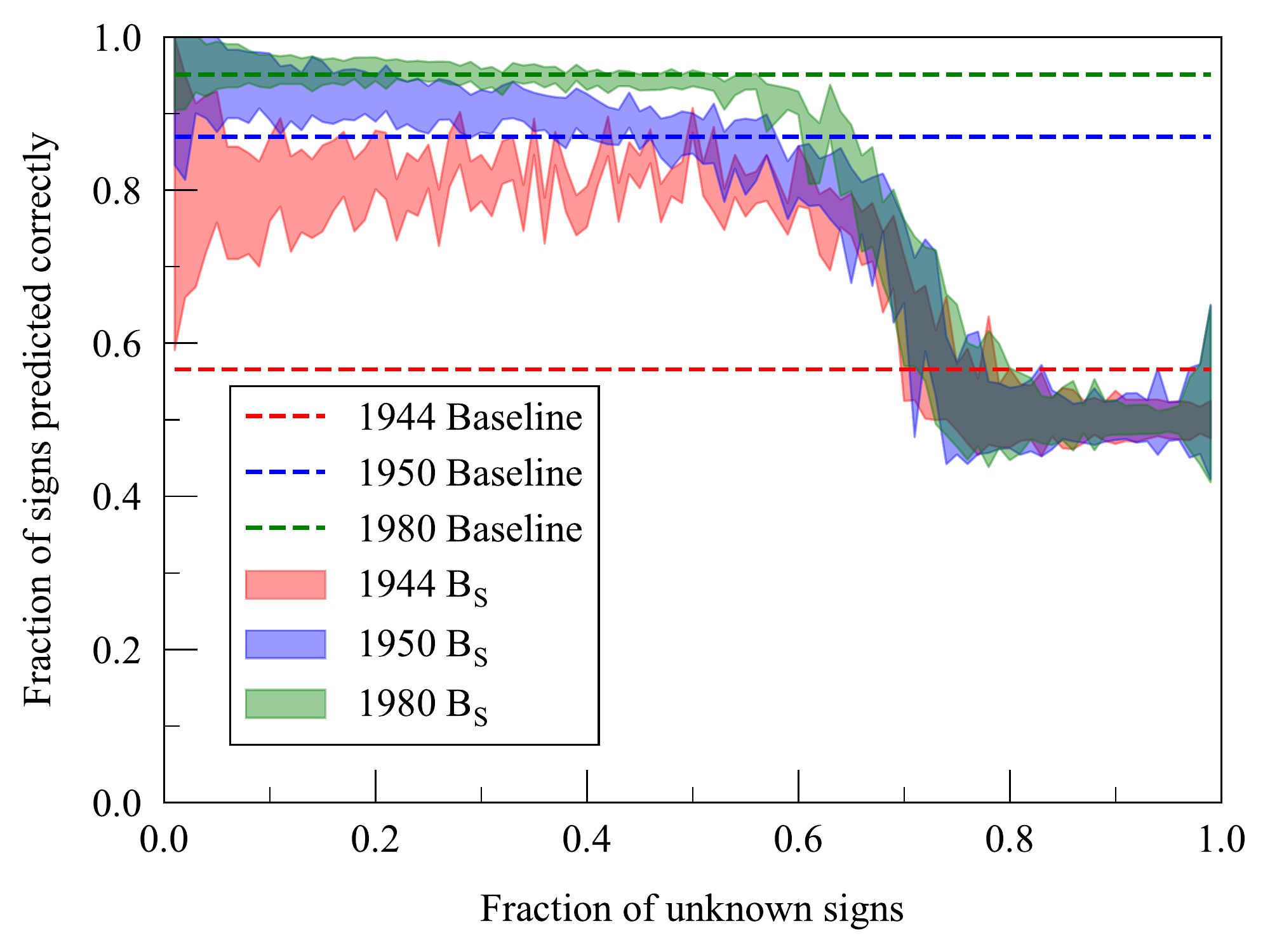}
\caption{Fraction of signs predicted correctly in the multiple sign prediction task using the strong balance measure~$B_S$, as a function of the fraction of unknown signs for the international relations networks in the years 1944, 1950, and 1980.}
\label{fig:strong_acc}
\end{figure}

\begin{figure}
\centering
\includegraphics[width=\columnwidth]{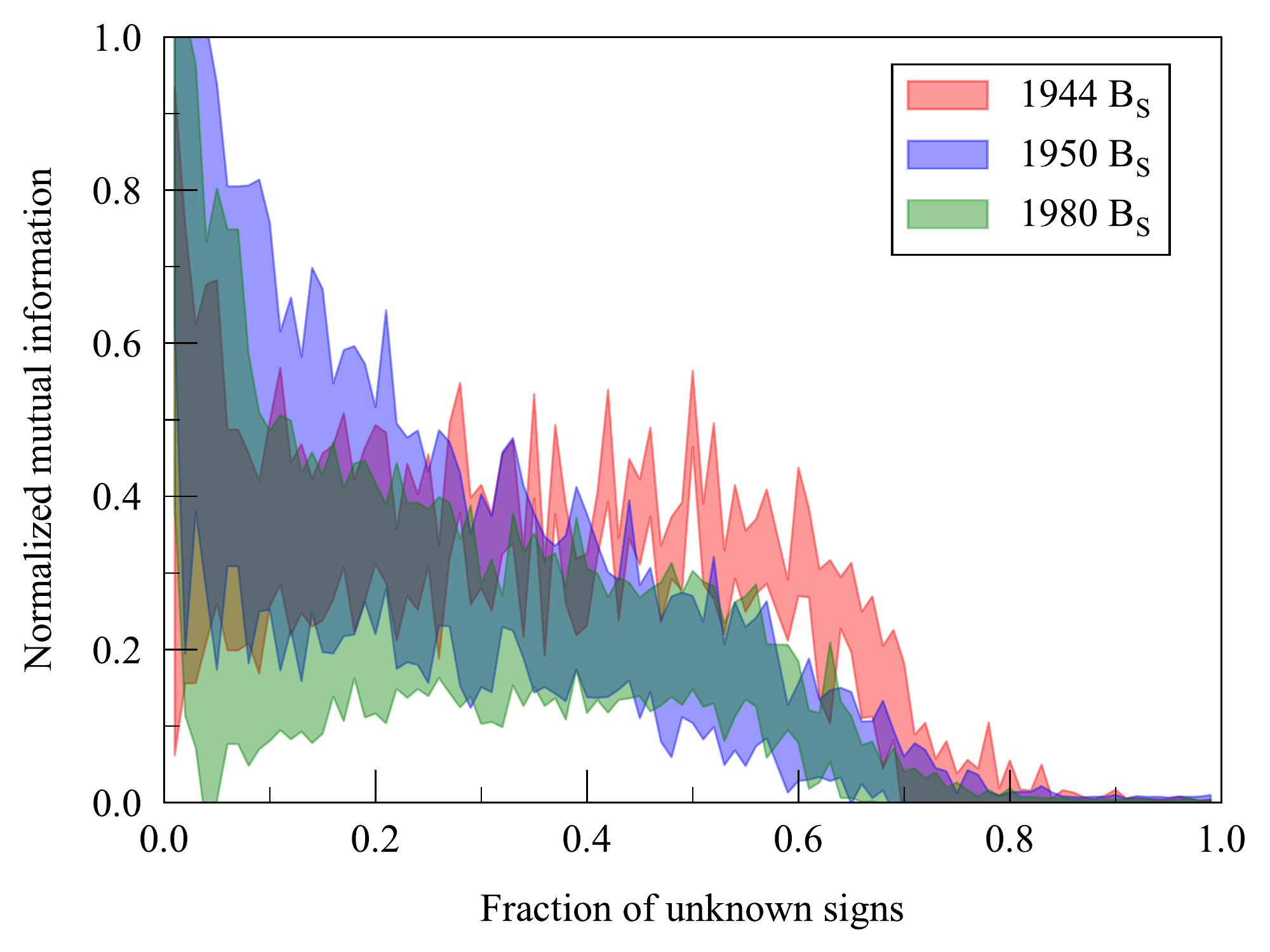}
\caption{Normalized mutual information for the multiple sign prediction task using the strong balance measure~$B_S$, as a function of the fraction of unknown signs for the international relations networks in the years 1944, 1950, and 1980.}
\label{fig:strong_nmi}
\end{figure}

Figures~\ref{fig:weak_acc}, \ref{fig:weak_nmi}, \ref{fig:strong_acc}, and~\ref{fig:strong_nmi} show accuracy and NMI results from calculations for three of our international relations networks, corresponding to the years 1944 (during the Second World War, where 43\% of signs are negative), 1950 (a few years afterward, where 13\% of signs are negative), and 1980 (relative peace, where 5\% of signs are negative).  Each plot shows three separate curves for the three networks, as a function of the fraction of signs removed from the network.  For the accuracy plots we also show the baselines set by assuming that all unknown signs are positive.  (For the NMI plots the equivalent baselines are by definition at zero.)

As the fraction of signs removed gets larger (and hence the amount of information remaining to learn from gets smaller) we naturally expect the performance of the algorithm to fall off.  Figures~\ref{fig:weak_acc} and~\ref{fig:weak_nmi} show results for the weak balance measure~$B_W$ and reveal that predictions are reasonably accurate for all three years studied for fractions of predicted signs up to about~50\%, although the baseline accuracy for 1980 is so high that it is comparable with the predictions.  (This is simply because a very large fraction of signs are positive in this network.)  Normalized mutual information is also well above the baseline level of zero for fractions of predicted signs up to about~50\%.  Beyond th 50\% mark, however, prediction accuracy rapidly falls to close to zero.

Figures~\ref{fig:strong_acc} and~\ref{fig:strong_nmi} show the corresponding results for the strong balance measure~$B_S$, and comparing the results for the two measures reveals an interesting overall picture.  The weak measure does better when predicting smaller numbers of signs, but suddenly fails around the 50\% mark, beyond which it does no better (in fact worse) than chance.  The strong measure, by contrast, does less well when fewer than 50\% of signs are removed, but manages at least modestly good performance well beyond the 50\% point, thereby outperforming the weak measure in this regime (although it is still not very good).  These trends are especially clear in the 1944 network, for which arguably the strong measure makes more sense since, as discussed earlier, international relations were dominated by two main factions during that era.

The failure of the weak balance metric to predict unknown signs beyond about the 50\% mark is particularly interesting.  It arises through a competition between two different minima of the metric.  One minimum approximately corresponds to the true assignment of signs, and if the algorithm finds this minimum it will succeed, at least partially, in the sign prediction task.  The other minimum is a trivial one in which all, or almost all, unknown signs are negative.  If the fraction of unknown signs is large enough, the latter state will contribute at least two negative signs to most closed loops in the network, meaning that most loops are balanced (according to the weak definition) and hence our imbalance score will approach its lowest possible value of zero.  As the fraction of unknown signs grows, there comes a point at which this trivial minimum outcompetes the nontrivial one and the algorithm no longer predicts signs with success any better than chance.  This point---the discontinuity we see in Fig.~\ref{fig:weak_nmi}---is in effect a zero-temperature first-order phase transition between competing ground states.  No similar argument applies to the strong balance measure, and hence we see no sharp phase transition in that case.

Overall, we conclude that successful prediction of multiple edge signs is possible using our balance measures, with the weak notion of balance again giving better performance than the strong notion, at least up to the phase transition mentioned above, beyond which the strong balance measure is a better choice.  In the particular networks examined here, performance is stronger for the years 1944 and 1950 than for 1980, perhaps because of the starker conflicts and alliances during and immediately after the war, compared with the relative peace of the early 1980s.

\section{Conclusions}
We have studied the phenomenon of structural balance in signed networks, whereby some configurations of signed edges are more common than others.  We have proposed two measures of structural balance based on previously hypothesized notions of ``weak'' and ``strong'' balance and compared their performance against each other and previously proposed measures in a number of tasks.  Specifically, we have examined the behavior of the various measures on two distinct sets of networks representing alliances and conflicts between countries during the 20th and 21st centuries, as well as university freshman cohort relationships, testing in the first instance to see simply by which measures these networks are most balanced.  We find that all measures show a significant level of balance in all of the networks we study.

We further test our measures on the international relations data by comparing their ability to predict unknown edge signs in a set of cross-validation experiments, in which we remove either a single sign or multiple signs from the network and attempt to predict the missing sign(s) by choosing those values that maximize balance by each of our metrics.  We find that prediction of unknown signs is possible, with accuracy substantially better than a random guess, and in particular that our measure based on the weak notion of balance performs well in practice.

Many extensions and generalizations of the work presented here would be possible.  Good data on signed networks are currently relatively scarce, but it would be interesting to see how our results generalize when similar calculations are performed on other networks. As discussed in Section~\ref{sec:balance}, many data sets are more naturally represented as weighted and/or directed signed networks, and so extending the measures proposed here to these classes of networks would provide a more flexible framework for analysis of a wide variety of data.  One could also employ balance metrics to perform anomaly detection in networks, looking for edges that participate in a large number of imbalanced loops.  A further interesting question is how to determine the optimal value of the parameter we called~$\alpha$, which controls the amount by which longer loops are discounted in our calculations.  In this paper we simply choose a value that seems reasonable, noting that our results are not strongly dependent on the choice, but it would be an improvement if one were able to find a first-principles method of fixing the value of~$\alpha$.  These possibilities, however, we leave for future work.

\begin{acknowledgments}
The authors thank Robert Axelrod, Maria Riolo, and Jean-Gabriel Young for useful discussions.  This work was funded in part by the US National Science Foundation under grants DMS--1407207 and DMS--1710848.
\end{acknowledgments}

\end{document}